\documentclass[useAMS,usenatbib]{mn2e}

\usepackage{epsfig}
\usepackage{amsmath}
\usepackage{amssymb}
\usepackage{natbib}
\usepackage{threeparttable} 
\usepackage{booktabs}

\usepackage{epstopdf}
\usepackage{times}

\newcommand{\swift}{\textit{Swift}}
\newcommand{\rxte}{\textit{RXTE}}

\newcommand{\maxi}{\textit{MAXI}}

\newcommand{\nustar}{\textit{NuSTAR}}

\newcommand{\Msun}{\mathrm{M}_{\odot}}
\newcommand{\lum}{\mathrm{erg~s}^{-1}}
\newcommand{\flux}{\mathrm{erg~cm}^{-2}~\mathrm{s}^{-1}}
\newcommand{\fluence}{\mathrm{erg~cm}^{-2}}
\newcommand{\cnts}{\mathrm{counts~s}^{-1}}

\newcommand{\mdotgs}{\mathrm{g~s}^{-1}}
\newcommand{\nh}{\mathrm{cm}^{-2}}

\newcommand{\dist}{(D/3.6~\mathrm{kpc})^2}

\newcommand{\gmc}{GM/c^2}

\newcommand{\source}{4U 1608--52}

\def \mnras {MNRAS}
\def \apj {ApJ}
\def \apjs {ApJS}

\def \apjl {ApJL}
\def \aap {A\&A}
\def \nat {Nature}

\def \atel {ATel}

\def \pasj {PASJ}

\def \ssr {Space Sci. Rev.}

\title[NuSTAR X-ray burst from 4U 1608--52]{Probing the effects of a thermonuclear X-ray burst on the neutron star accretion flow with \nustar}
\author[N. Degenaar et al.]
{N. Degenaar$^1$\thanks{e-mail: degenaar@ast.cam.ac.uk}, K.~I.~I.~Koljonen$^2$, D.~Chakrabarty$^{3}$, E.~Kara$^1$, D.~Altamirano$^4$,  
\newauthor J.~M.~Miller$^{5}$, and A.~C.~Fabian$^1$\\
$^1$Institute of Astronomy, University of Cambridge, Madingley Road, Cambridge CB3 OHA, UK\\
$^2$New York University Abu Dhabi, PO Box 239188 Abu Dhabi, UAE\\
$^3$Massachusetts Institute of Technology (MIT), Kavli Institute for Astrophysics and Space Research, Cambridge, MA 02139, USA\\
$^4$Department of Physics and Astronomy, University of Southampton, Southampton, Hampshire, SO171BJ, UK\\
$^5$Department of Astronomy, University of Michigan, 1085 South University Avenue, Ann Arbor, MI  48109, USA
}

\begin{document}

\date{Accepted 2015 December 16.  Received 2015 December 9; in original form 2015 November 10}

\pagerange{\pageref{firstpage}--\pageref{lastpage}} \pubyear{0000}

\maketitle

\label{firstpage}

\begin{abstract}
Observational evidence has been accumulating that thermonuclear X-ray bursts ignited on the surface of neutron stars influence the surrounding accretion flow. Here, we exploit the excellent sensitivity of \nustar\ up to 79~keV to analyse the impact of an X-ray burst on the accretion emission of the neutron star LMXB \source. The $\simeq$200~s long X-ray burst occurred during a hard X-ray spectral state, and had a peak intensity of $\simeq$30--50 per cent of the Eddington limit with no signs of photospheric radius expansion. Spectral analysis suggests that the accretion emission was enhanced up to a factor of $\simeq$5 during the X-ray burst. We also applied a linear unsupervised decomposition method, namely non-negative matrix factorisation (NMF), to study this X-ray burst. We find that the NMF performs well in characterising the evolution of the burst emission and is a promising technique to study changes in the underlying accretion emission in more detail than is possible through conventional spectral fitting. For the burst of \source, the NMF suggests a possible softening of the accretion spectrum during the X-ray burst, which could potentially be ascribed to cooling of a corona. Finally, we report a small ($\simeq$3 per cent) but significant rise in the accretion emission $\simeq$0.5~h before the X-ray burst, although it is unclear whether this was related to the X-ray burst ignition.
\end{abstract}


%%%%%%%%%%%
% INTRODUCTION
%%%%%%%%%%%

\section{Introduction}\label{sec:intro}
In low-mass X-ray binaries (LMXBs) a black hole or neutron star accretes matter from a (sub-) solar-mass companion star. The donor typically overflows its Roche lobe and transfers mass to an accretion disc that surrounds the compact primary. Thermal-viscous instabilities in the accretion disc are thought to be the reason that many LMXBs are transient, i.e., cycle through outburst and quiescent episodes \citep[e.g.,][for a review]{lasota01}. Material is rapidly accreted during weeks--months long outbursts, resulting in X-ray luminosity of $L_{\mathrm{X}}$$\simeq$$10^{35}-10^{39}~\lum$. In intervening quiescent phases, the luminosity is orders of magnitude lower and little or no matter is reaching the black hole or neutron star. 

Along the course of an accretion outburst, prominent changes in the X-ray spectral and variability properties can be observed that are thought to reflect changes in the accretion geometry \citep[e.g.,][]{homan2005,vanderklis2006review,gladstone2007,munozdarias2014}. Broadly, the distinction is made between soft and hard X-ray spectral states, depending on the relative contributions of the soft X-rays emitted by the accretion disc and the hard X-rays produced by a corona (a population of hot electrons, the exact nature of which is not established) to the overall X-ray spectrum. Moreover, hard states show stronger variability, with lower characteristic frequencies, than soft states.

In the case of neutron stars, matter accumulates on the stellar surface where it undergoes thermonuclear burning that converts the accreted H/He into heavier elements \citep[e.g.,][for reviews]{lewin95,strohmayer06,schatz2006}. If radiative cooling of the burning layer is slower than the energy generation rate, a thermonuclear runaway occurs that causes the entire layer to ignite on a time-scale of $\simeq$1~s. This explosive energy release results in a bright burst of X-ray emission characterised by an initial fast rise due to the rapid burning of the fuel layer, followed by a slower decay phase as the burning ashes cool. Thousands of such type-I X-ray bursts (shortly X-ray bursts hereafter) have been observed from $\simeq$100 Galactic neutron star LMXBs \citep[e.g.,][]{cornelisse2003,galloway06}.

The spectra of X-ray bursts are blackbody-like with a temperature of $kT_{\mathrm{bb}}$$\simeq$1--3~keV that evolves over time. At the peak, the X-ray burst emission may exceed the Eddington limit ($L_{\mathrm{Edd}}\simeq2\times10^{38}~\lum$), causing the photosphere of the neutron star to expand due to radiation pressure. Such photospheric radius expansion (PRE) can be identified through time-resolved spectroscopy as a peak in the blackbody emission area that is accompanied by a dip in the observed temperature \citep[e.g.,][]{lewin1993}.

It has long been realised that there is an intricate connection between the properties of the accretion flow and the observable manifestation of X-ray bursts. For example, the rate at which mass is accreted on to the neutron star determines the duration and recurrence time of X-ray bursts \citep[e.g.,][]{fujimoto81,bildsten98,cornelisse2003,peng2007}. Furthermore, the accretion geometry  influences the ignition conditions \citep[e.g.,][]{cavecchi2011,zand2012}, spectral evolution \citep[e.g.,][]{kajava2014,poutanen2014_4u1608}, and rapid variability \citep[e.g.,][]{muno2004,galloway06} of X-ray bursts. 

In turn, the sudden and intense radiation caused by X-ray bursts can affect the surrounding accretion flow e.g., by cooling or ejecting the corona \citep[e.g.,][]{maccarone2003,kluzniak2013,mishra2014}, and exerting a drag on the accretion disc, driving a mass outflow from it, or changing its structure \citep[e.g.,][]{walker1989,miller1996,ballantyne2005}. Indeed, observational evidence has been accumulating that all these various effects occur. For example, hard (30--50 keV) flux deficits seen during X-ray bursts of several sources may be indicative of coronal cooling \citep[][]{maccarone2003,chen2012_xrbs_igrj1747,chen2013_aqlx1_bursts,ji2013_xrbs_4u1636,ji2014_bursts,ji2014_aa}. Furthermore, absorption and emission features detected during particularly long and intense X-ray bursts may point to a mass outflow or sudden changes in the accretion disc structure \citep[e.g.,][]{ballantyne2004,ballantyne2005,zand2011,degenaar2013_igrj1706,keek2014_refl}. Moreover, several recent studies suggest that the accretion emission is temporarily enhanced during X-ray bursts, possibly resulting from a radiation drag asserted on the inner accretion disc \citep[][]{intzand2013,worpel2013,worpel2015,ji2014_4u1608,keek2014,peille2014}. 

All of the above physical processes are likely operating, possibly competing, during X-ray bursts. Given that hard and soft X-ray states are thought to reflect different accretion geometries, the impact of X-ray bursts is likely state dependent. Indeed, a reduction of the $>$30~keV emission is only seen for X-ray bursts occurring in the hard X-ray spectral state when the corona contributes most prominently to the overall X-ray spectrum \citep[][]{chen2012_xrbs_igrj1747,chen2013_aqlx1_bursts}. Furthermore, the enhancement of the accretion emission shows different patterns in the two different spectral states \citep[e.g.,][]{ji2014_4u1608}. 

Studying accretion changes during an X-ray burst is somewhat challenging for soft X-ray states, because the accretion emission is then difficult to disentangle from the soft X-ray burst radiation \citep[e.g.,][]{barriere2014_grs,worpel2015}. On the other hand, during hard X-ray states both the accretion flux and the X-ray burst flux are typically lower, and hence good sensitivity is required to discern how a burst affects the accretion flow. In this work we exploit the excellent sensitivity of \nustar\ up to an energy of $\simeq$79 keV, to study an X-ray burst from the neutron star LMXB \source, detected during a hard X-ray state in 2014 \citep[][]{degenaar2015_4u1608}.

\begin{figure}
 \begin{center}
\includegraphics[width=8.3cm]{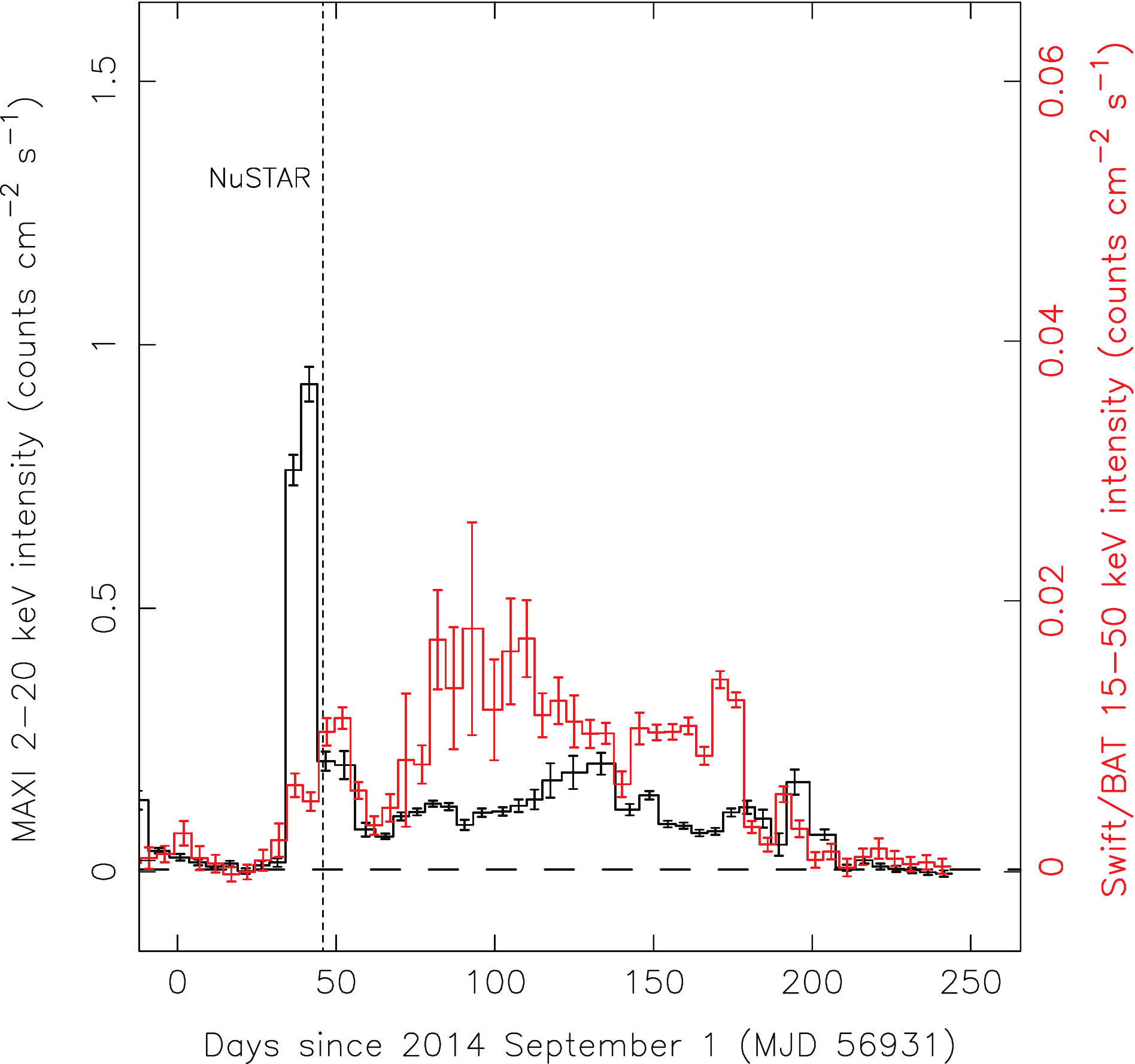}
    \end{center}
\caption[]{{\maxi/GSC (2--20 keV, black) and \swift/BAT (15--50 keV, red) light curves of the outburst that started in 2014 October (5-d bins). The time of our \nustar\ observation is indicated by the vertical dotted line. \\}}
 \label{fig:maxibat}
\end{figure} 

\begin{figure*}
 \begin{center}
\includegraphics[width=8.0cm]{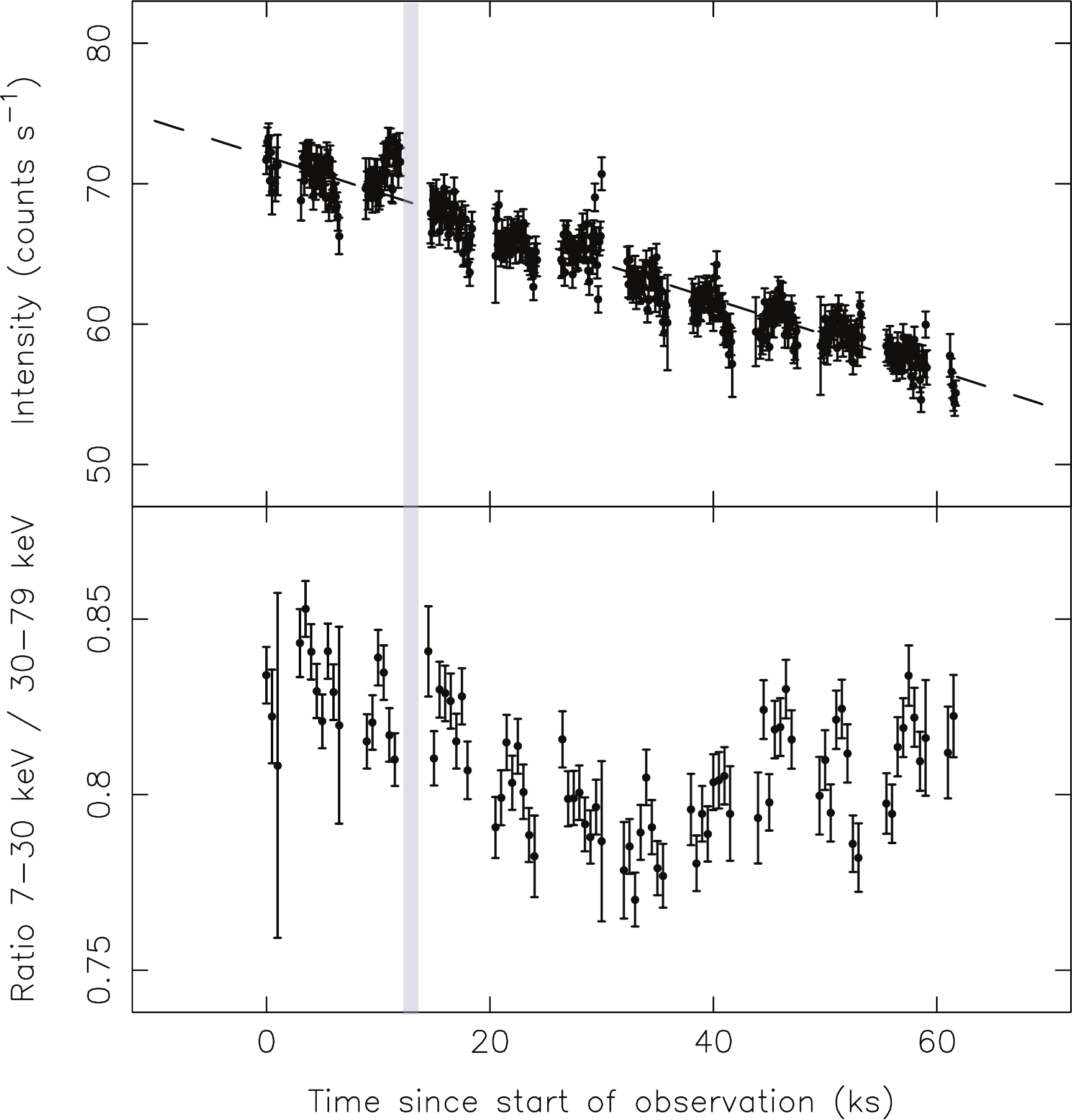}\hspace{+0.2cm} 
\includegraphics[width=8.0cm]{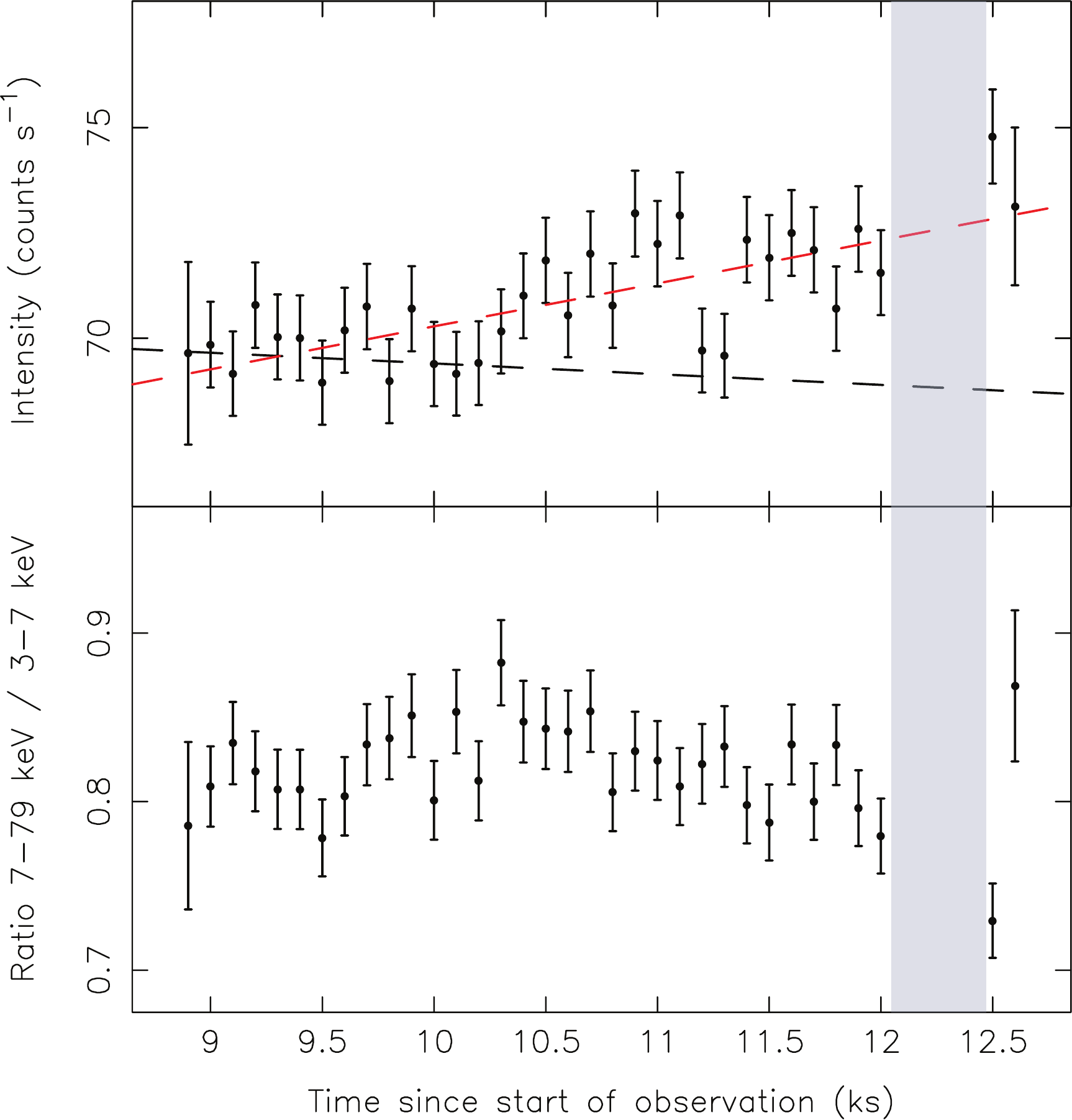} 
    \end{center}
\caption[]{{\nustar\ FPMA/FPMB summed, background-corrected 3--79 keV count rate light curve (top panels) and 7--30 keV / 3--7 keV hardness ratio (bottom panels). The X-ray burst was cut out for representation purposes and is shown separately in Fig.~\ref{fig:lcE}; the time at which it occurred is indicated by the vertical grey bars. In the top panels the dashed black line indicates a linear decay fit to the overall light curve (i.e., using all satellite orbits). Left: the entire observation using 100 s bins for the light curve and 500 s bins for the hardness ratio. Right: zoom of \nustar\ orbit 3, using 100 s bins for both the light curve and the hardness ratio. During this orbit the count rate evolution was better described by a linear function with a positive slope (red dashed line).
}}
 \label{fig:lc}
\end{figure*}

\subsection{\source}
\source\ is a prolific transient neutron star LMXB and X-ray burster. Since its discovery in 1972 \citep[][]{grindlay1976,tananbaum1976}, accretion outbursts have been observed roughly once every 1--2 years \citep[e.g.,][]{lochner1994,chen97,simon2004,galloway06}. Occasionally these outbursts are bright, reaching up to an X-ray luminosity of $L_{\mathrm{X}}$$\simeq$$(1-5) \times 10^{37}~\dist~\lum$. However, more often the source exhibits fainter outbursts that peak at few times $10^{36}~\dist~\lum$. 

In addition to regular X-ray bursts, one superburst was detected from \source: A very long and energetic explosion thought to result from the ignition of a thick layer of carbon \citep[][]{keek2008}. Assuming that its brightest X-ray bursts reach the Eddington limit, a source distance of $D$$\simeq$2.9--4.5~kpc has been estimated \citep[e.g.,][]{galloway06,poutanen2014_4u1608}.\footnote{An alternative method based on interstellar extinction suggests a larger distance of $D=5.9^{+2.0}_{-1.9}$~kpc \citep[][]{guver2010}, although the X-ray burst behaviour favours a closer distance \citep[][]{poutanen2014_4u1608}.} The detection of burst oscillations has revealed the spin period of the neutron star \citep[$P_{\mathrm{s}}$$\simeq$620~Hz;][]{muno2001,galloway06}. 

In 2014 October, \maxi\ detected renewed activity from \source\ \citep[][see Fig.~\ref{fig:maxibat}]{negoro2014_4u1608}. An observation with \nustar\ \citep[][]{harrison2013_nustar} performed shortly after the onset of this outburst revealed the characteristic signatures of disc reflection: a broad Fe-K line peaking near 7 keV and a Compton hump at $\simeq$20--30 keV \citep[][]{degenaar2015_4u1608}. The high data quality allowed us to constrain the accretion geometry in terms of the inner disc radius ($R_{\mathrm{in}}$$\simeq$7--10~$\gmc$), disc inclination ($i$$\simeq$$30^{\circ}$--$40^{\circ}$) and height of the corona ($h$$<$8.5$~\gmc$). The measured 3--79 keV unabsorbed flux was $F_{3-79}\simeq$$2\times10^{-9}~\flux$, of which $>$95 per cent could be ascribed to a hard spectral component modelled by a $\Gamma \simeq 2$ power law with a high-energy cutoff of $E_{\mathrm{cut}}\gtrsim$300~keV. This spectral shape strongly suggests that \source\ was detected in a hard X-ray state during the \nustar\ observation. Furthermore, it was accreting at a few per cent of the Eddington limit \citep[][]{degenaar2015_4u1608}. In this work we report on the detailed analysis of the X-ray burst that was detected during the \nustar\ observation of \source.

%%%%%%%%%%%
% OBS+ANA
%%%%%%%%%%%

\section{Observations and Analysis}
Fig.~\ref{fig:maxibat} shows the monitoring light curves of \source\ obtained with \maxi/GSC \citep[2--20 keV;][]{maxi2009} and \swift/BAT \citep[15--50 keV;][]{krimm2013}. We observed the source with \nustar\ during the early part of the $\simeq$5-month long outburst (Fig.~\ref{fig:maxibat}), on 2014 October 16--17 from 23:00 to 17:10 UT (Obs ID 90002002002). \nustar\ is sensitive at energies of 3--79 keV and consists of two co-aligned imaging telescopes, Focal Plane Mirror (FPM) A and B. The observation was spread over 12 different satellite orbits within a time frame of $\simeq 60$~ks, as illustrated in Fig.~\ref{fig:lc} (top left). 

We reduced and analysed the \nustar\ data using tools incorporated in \textsc{heasoft} version 16.6. Standard screening with \textsc{nustardas} resulted in a net on-source exposure time of $\simeq$32~ks. We created light curves and spectra with the \textsc{nuproducts} tool, using an aperture of 120 arcsec for the source and one of similar size for the background. The FPMA/FPMB light curves were background-corrected and summed using \textsc{lcmath}. \source\ was significantly detected above the background in the entire \nustar\ band. 

Since the spectra obtained with the two individual mirrors showed excellent agreement, we opted to combine the spectral data and ancillary response matrix files using \textsc{addascaspec}. The redistribution matrix files were combined with \textsc{addrmf}, where we weighted the responses of the two mirrors by their exposure times. All spectra were grouped to a minimum of 20 counts per bin and fitted using \textsc{XSpec} \citep[][]{xspec}. Throughout this work we use a distance of $D=3.6$~kpc towards \source\ and report errors as 1$\sigma$ confidence intervals unless noted otherwise.

%%%%%%%%%%%
% RESULTS
%%%%%%%%%%%

\section{Results}\label{sec:results}

\subsection{Light curve analysis}\label{subsec:lc}

\subsubsection{Persistent accretion emission}\label{subsec:lcacc}
Fig.~\ref{fig:lc} shows the evolution of the persistent emission along the $\simeq$60-ks window during which \source\ was observed with \nustar. An X-ray burst was detected during the third satellite orbit (grey shaded area in Fig.~\ref{fig:lc}), which is shown in Fig.~\ref{fig:lcE}. During the observation, the FPMA/FPMB summed count rate of the persistent (i.e., non-burst) accretion emission decreased by $\simeq$25 per cent from $\simeq$75$~\cnts$ at the start of the observation to $\simeq$55$~\cnts$ at the end. Making energy cuts shows that the decay is visible for energies up to $\simeq$30 keV; the count rate in the 30--79 keV band is consistent with being constant along the observation. Defining a hardness ratio as the ratio of counts in the 7--30 and 3--7 keV ranges suggests that there are no prominent spectral changes associated with the decay in count rate (Fig.~\ref{fig:lc} bottom left).

\begin{figure}
 \begin{center}
\includegraphics[width=8.0cm]{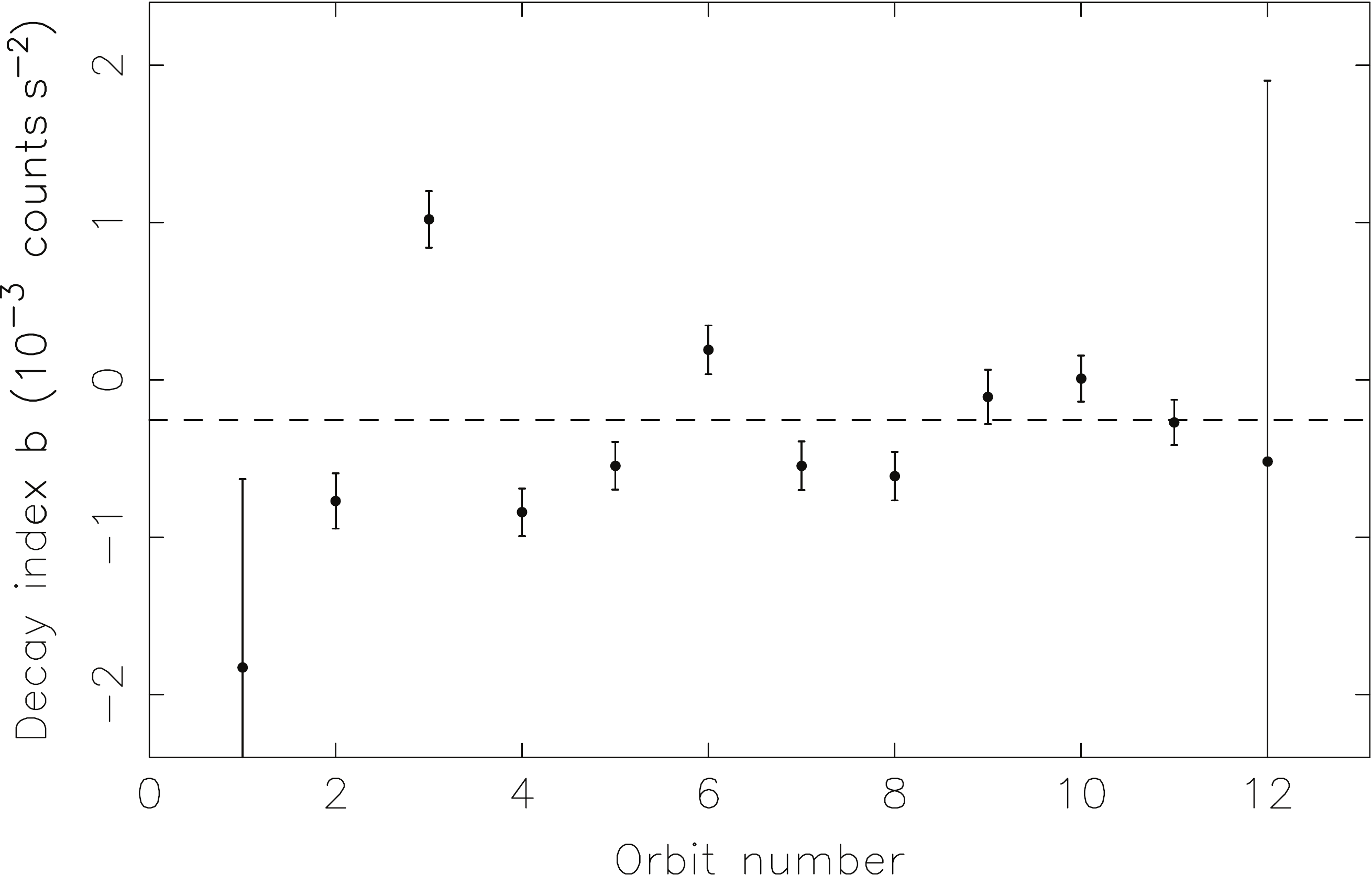}\hspace{+0.2cm} 
    \end{center}
\caption[]{{Values for the index of a linear fit to the count rate light curve for each individual satellite orbit, where $b$$<$0 indicates a decreasing intensity. The average value obtained from fitting the entire \nustar\ light curve (i.e., all 12 orbits; Fig.~\ref{fig:lc}) is indicated by the dashed line. The X-ray burst occurred during \nustar\ orbit 3, and was cut from the data for these fits. Errors are 1$\sigma$.
}}
 \label{fig:linfit}
\end{figure}

The overall light curve, excluding the X-ray burst, can be reasonably described by a simple linear function of the form $f(t)$$=$$a$$+$$bt$ where $t$ is the time since the start of the \nustar\ observation. We obtain a constant offset of $a$$=$$71.89 \pm 0.11~\cnts$ and a slope of $b$$=$$-(2.55 \pm 0.03)\times10^{-4}$~counts~s$^{-2}$ ($\chi_{\nu}^2$=1.50 for 749 dof). This fit is shown as the dashed line in Fig.~\ref{fig:lc}. Individual satellite orbits could be described by decay slopes lying within $\simeq$3$\sigma$ of the value obtained for the overall light curve, as shown in Fig.~\ref{fig:linfit}. The exception is orbit 3, which is best fit with a positive slope of $b$$=$$(1.02 \pm 0.18)\times10^{-3}$~counts~s$^{-2}$ ($a$$=$$60.08 \pm 1.93~\cnts$, $\chi_{\nu}^2$=0.73 for 65 dof) and lies $\simeq$7$\sigma$ from the overall value (Fig.~\ref{fig:linfit}). Fig.~\ref{fig:lc} (top right) shows that the count rate started to rise $\simeq$1~ks into the observation, which was $\simeq$1.5~ks ($\simeq$0.5~h) before the X-ray burst detection (it is unclear whether there is a connection; see Section~\ref{subsec:persenhance}). The count rate increased by $\simeq$3 per cent and there appears no change in hardness associated with it (Fig.~\ref{fig:lc}, bottom right). 

\subsubsection{Burst emission}\label{subsec:lcburst}
\begin{figure}
 \begin{center}
\includegraphics[width=8.3cm]{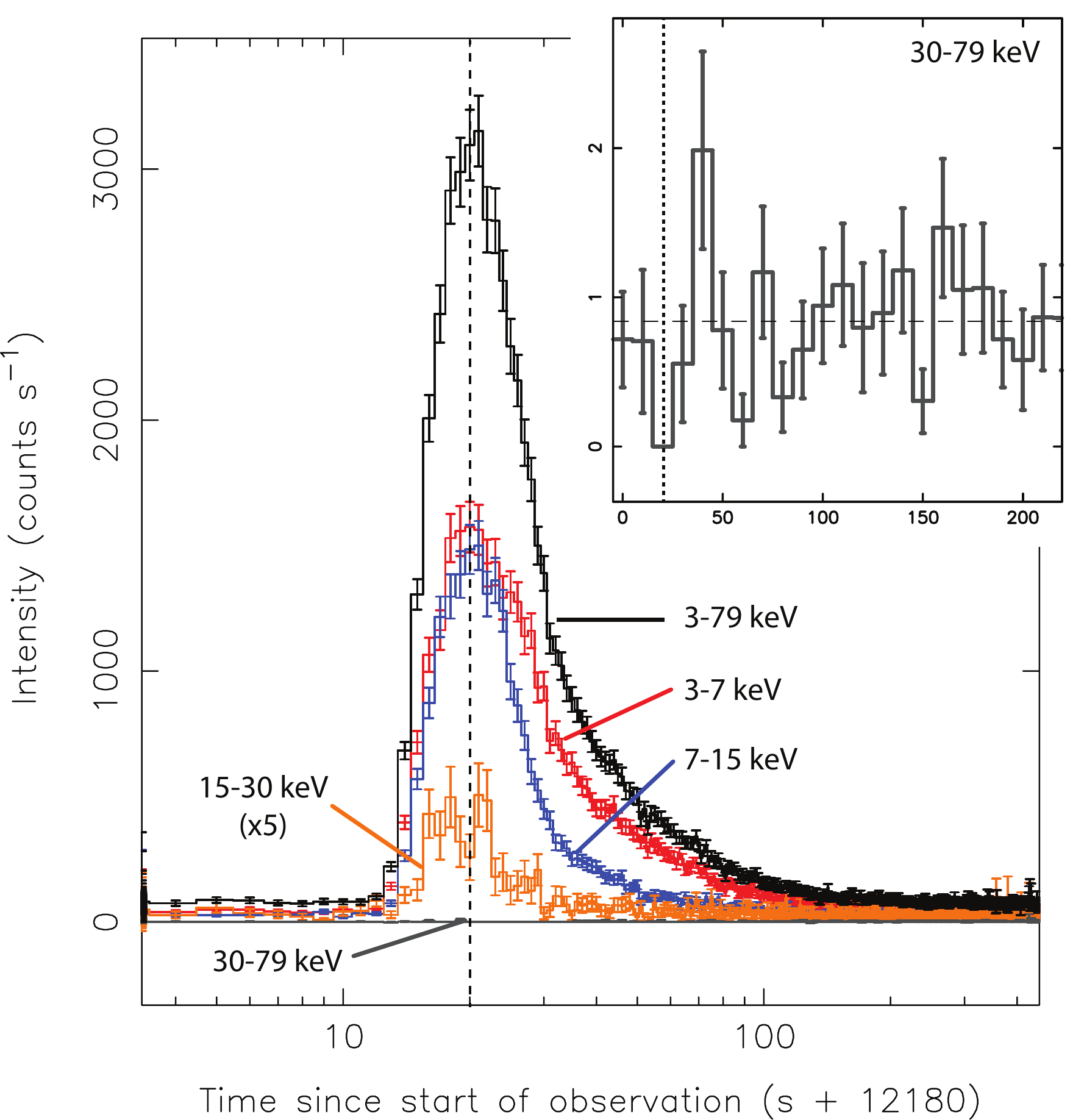}\hspace{+0.2cm} 
    \end{center}
\caption[]{{\nustar\ FPMA/FPMB summed, background-corrected count rate light curve of the X-ray burst in different energy bands (1-s bins). The inset shows the 30--79 keV light curve at 10-s resolution. The 15--30 keV count rates were multiplied by a factor of 5 for visual clarity.} The vertical dotted line marks the time of the X-ray burst peak and the dashed horizontal line the average 3--79 keV count rate over the entire observation.}
 \label{fig:lcE}
\end{figure}

Fig.~\ref{fig:lcE} shows the light curve of the X-ray burst in different energy bands at 1-s resolution. The burst was detected up to $\simeq$30 keV. The inset in Fig.~\ref{fig:lcE} shows a zoom of the 30--79 count rate light curve at 10-s resolution. The 30--79 keV count rate is zero during a single bin around the X-ray burst peak, and there seems to be a small reduction in the 15--30 keV count rate at this time (dotted vertical line). However, the error bars are large and this possible `dip' is not significant. 

The pre-burst 3--79 keV count rate was $\simeq$70~$\cnts$ (Fig.~\ref{fig:lc}) and the source returned to this level $\simeq$210~s after the start of the X-ray burst. We note that the duration of X-ray bursts depends strongly on the energy band considered. This is illustrated by Fig.~\ref{fig:lcE}, which shows that the X-ray burst decayed more rapidly at higher energies. This behaviour is typical and results from the gradual decrease in temperature as the burning ashes cool along the burst decay. The X-ray burst peaked $\simeq$10~s after the initial rise.

\subsection{Time-resolved spectroscopy of the X-ray burst}\label{subsec:spec}
To investigate the spectral evolution during the X-ray burst, we selected 14 consecutive intervals of varying length ($\simeq$2--65~s) to ensure $\simeq$1800~counts per spectrum. For all burst intervals, a spectrum created from $\simeq$100~s of data preceding the burst was subtracted as the underlying accretion emission. To model the burst we used a simple blackbody model (\textsc{bbodyrad}) and to account for interstellar extinction we employed the \textsc{tbabs} model with the abundances set to \textsc{wilm} \citep[][]{wilms2000}, and the cross-sections to \textsc{vern} \citep[][]{verner1996}. Once a best fit was found, we calculated the bolometric flux using the \textsc{cflux} convolution model with the energy boundaries set to 0.01 and 100 keV. 
Due to the energy cut-off at 3~keV, the hydrogen column density is not well constrained by the \nustar\ data. We therefore fixed it to $N_{\mathrm{H}}$$=$$1.5\times10^{22}~\nh$, which is within the range of values reported for \source\ \citep[$N_{\mathrm{H}}\simeq$$(1-2)\times10^{22}~\nh$; e.g.,][]{penninx1989,keek2008_1608,guver2010,degenaar2015_4u1608}. 

The black histograms in Fig.~\ref{fig:resolvedspec} show the results of our time-resolved spectroscopy. The bolometric flux peaks $\simeq$10~s after its initial rise, the measured blackbody temperatures are $kT_{\mathrm{bb}}$$\simeq$1.2--2.2~keV, and the  emission radii inferred from the blackbody normalisation are $R_{\mathrm{bb}}$$\simeq$4--7~km (for $D$$=$3.6~kpc). From our time-resolved spectroscopy we estimate a total fluence of $f_{\mathrm{bol}}$$\simeq$$1\times10^{-6}~\fluence$. Half of this is radiated during the first $\simeq$20~s of the burst. For $D$$=$3.6~kpc, the total radiated energy is $E_{\mathrm{burst}}$$\simeq$$1\times10^{38}$~erg. We find that the burst reached a peak flux of $F_{\mathrm{bol}}$$\simeq$$5.6\times10^{-8}~\flux$, which is a factor of $\simeq$15 higher than the pre-burst accretion emission. The Eddington flux of \source\ is not well measured: PRE bursts show a spread in bolometric peak flux of $F_{\mathrm{bol}}$$\simeq$(1.0--1.9)$\times10^{-7}~\flux$, which appears to depend on the accretion state \citep[e.g.,][]{kajava2014,poutanen2014_4u1608}. This suggests that the \nustar-detected X-ray burst peaked at $\simeq$30--50 per cent of the Eddington limit. The main burst properties are summarised in Table~\ref{tab:burst}.

\subsubsection{Allowing the persistent emission to vary during the burst}
In the standard approach outlined above we implicitly assume that the accretion emission is constant during an X-ray burst, but recent studies suggest that this might not be true \citep[e.g.,][]{intzand2013,worpel2013,worpel2015,ji2014_4u1608,keek2014}. We therefore repeated our spectral analysis by only subtracting the instrument background and then modelling the persistent emission together with the burst, as proposed by \citet{worpel2013}. 

Analysis of the time-averaged accretion spectrum revealed a power-law like shape with no evidence for a spectral cutoff in the \nustar\ bandpass \citep[][]{degenaar2015_4u1608}. We therefore use the \textsc{pegpwrlw} model for the accretion emission, the normalisation of which gives the unabsorbed model flux in a specified energy band (we choose 3--79 keV). A spectrum extracted from $\simeq$100~s prior to the X-ray burst can be described by $\Gamma$$=$$2.16\pm0.04$, and an unabsorbed 3--79 keV flux of $F_{3-79}$$=$$(2.16\pm0.07)\times10^{-9}~\flux$ (for $N_\mathrm{H}$$=$$1.5\times10^{22}~\nh$ fixed; $\chi_{\nu}^{2}=1.12$ for 104 dof). We fixed the \textsc{pegpwrlw} parameters to these values and then modelled the burst plus accretion emission as \textsc{tbabs*(bbodyrad+constant*pegpwrlw)}, where the constant factor ($f_a$) is introduced to parametrize changes in the accretion flux. The accretion emission was thus only allowed to change in normalisation, not in shape. Leaving the power-law index free resulted in very large errors, which indicates that we cannot address any possible changes in the shape of the spectral emission.

\begin{figure}
 \begin{center}
\includegraphics[width=8.9cm]{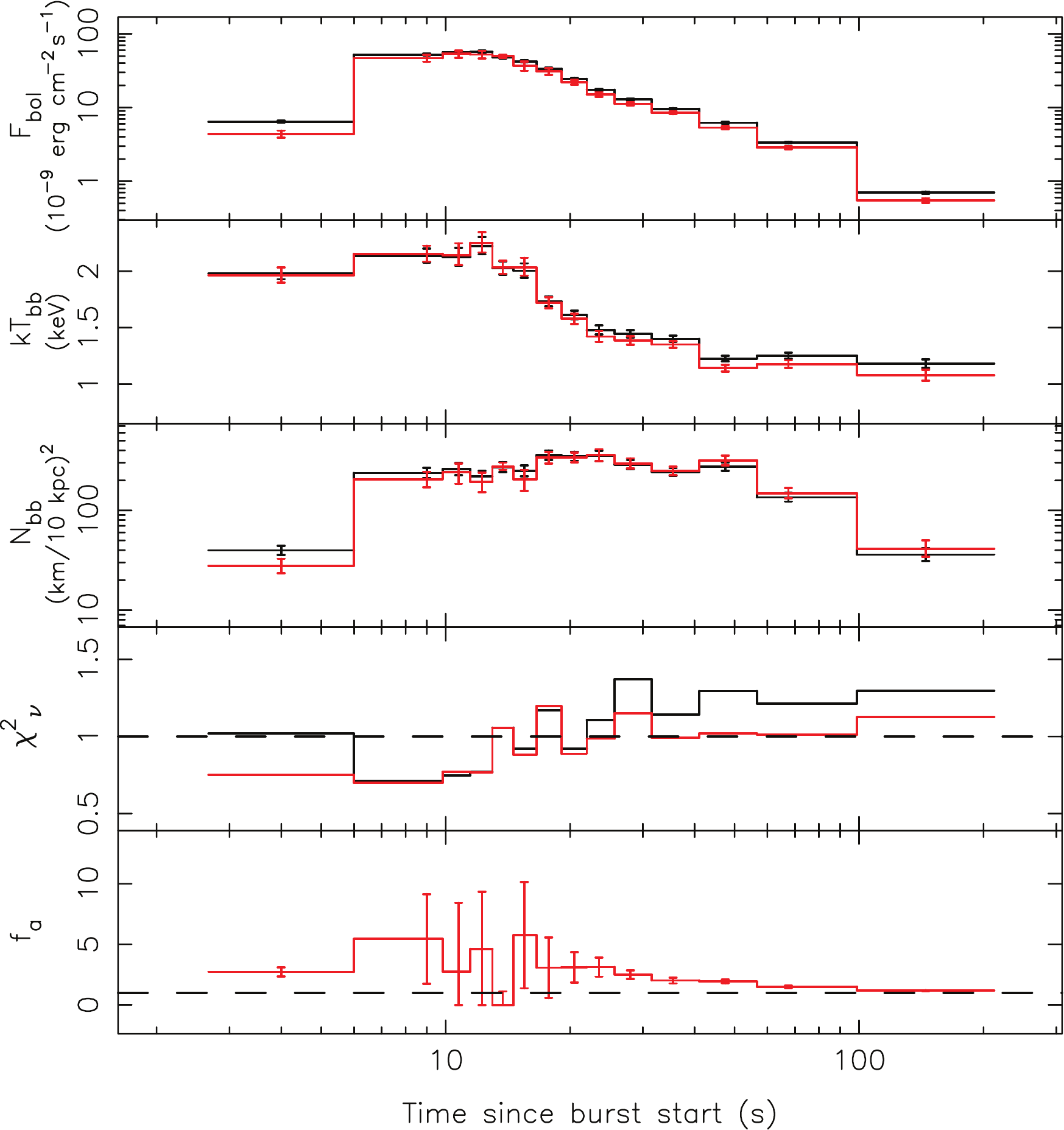} 
    \end{center}
\caption[]{{Results from time-resolved spectroscopy of the X-ray burst. In black the standard analysis in which a pre-burst persistent emission spectrum is subtracted from the burst spectra. In red an alternative method in which the persistent flux was allowed to vary during the burst by a factor $f_{a}$. From top to bottom: the bolometric thermal flux (0.01--100 keV), the blackbody temperature and normalisation, the reduced chi-square value of the fits, and the constant multiplication factor $f_{a}$. 
}}
 \label{fig:resolvedspec}
\end{figure}

\begin{figure}
 \begin{center}
\includegraphics[width=8.0cm]{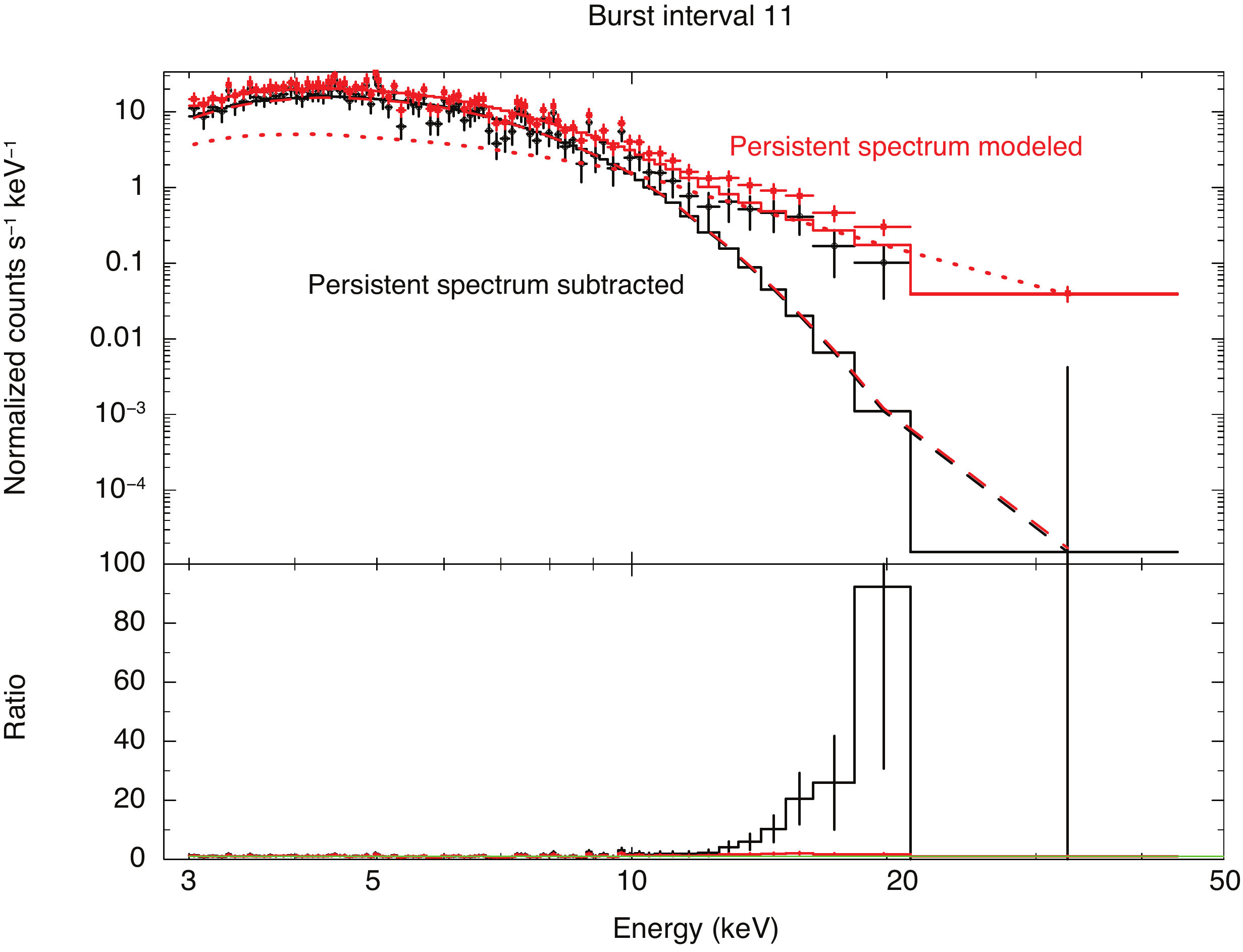} 
    \end{center}
\caption[]{{\nustar\ FPMA/FPMB combined spectrum and model fit (top panel), and data to model ratio (bottom panel) of an interval during the X-ray burst tail. In black the standard method where a spectrum of the pre-burst accretion emission is subtracted for the burst analysis. In red an alternative approach in which only the instrumental background is subtracted and the persistent emission is modelled to allow for changes in its normalisation. Dashed curves indicate the burst (blackbody) emission, and the dotted curve the accretion (power law) emission.}}
 \label{fig:spec}
\end{figure}

The results of allowing the persistent emission to change are shown as the red histograms in Fig.~\ref{fig:resolvedspec}. Overall, this provides a better fit to the burst spectra than keeping it fixed to the pre-burst value (fourth panel in Fig.~\ref{fig:resolvedspec}). This is also illustrated by Fig.~\ref{fig:spec}, which compares a spectrum taken $\simeq$35~s along the burst tail and fitted using the two different methods. When subtracting a pre-burst spectrum and fitting a blackbody to the burst (black curve in Fig.~\ref{fig:spec}), the model underestimates the data at $\gtrsim$15~keV. These residuals disappear when the persistent emission is allowed to vary in normalisation (red curve in Fig.~\ref{fig:spec}). We find that the constant multiplication factor is $f_{\mathrm{a}}$$>$1 during the larger part of the X-ray burst (bottom panel in Fig.~\ref{fig:resolvedspec}). It reached a maximum of $f_{\mathrm{a}}$$\simeq$5 around the burst peak, after which it gradually decreased and settled at $f_{\mathrm{a}}$$\simeq$1 around $\simeq$100~s after the start of the burst. This suggests that the accretion emission was temporarily enhanced in response to the X-ray burst.

\begin{table}
\caption{X-ray burst properties.}
\begin{threeparttable}
\begin{tabular*}{0.49\textwidth}{@{\extracolsep{\fill}}lc}
\hline
Parameter (unit)  & Value \\
\hline
Total duration, $t_{\mathrm{b}}$ (s)  & $\simeq 210$  \\
Total fluence, $f_{\mathrm{tot}}$ ($\fluence$)  & $\simeq 1.0 \times 10^{-6}$   \\
Radiated energy, $E_{\mathrm{b}}$ (erg)  & $\simeq 1.2 \times 10^{38}$   \\
Bolometric peak flux, $F_{\mathrm{bol}}$ ($\flux$)  & $\simeq 5.6 \times 10^{-8}$   \\
Accretion flux, $F_{\mathrm{acc}}$ ($\flux$)  & $\simeq 4 \times 10^{-9}$ \\
\hline
\end{tabular*}
\label{tab:burst}
\begin{tablenotes}
\item[] {\it Notes.} The bolometric (unabsorbed) peak flux is for the 0.01--100 keV energy range. A distance of $D=3.6$~kpc was assumed to calculate the total energy radiated during the X-ray burst. For reference, we note that the bolometric peak flux measured during PRE bursts of \source\ suggests that the Eddington flux lies in a range of $F_{\mathrm{Edd}}$$\simeq$$(1.0-1.9)\times10^{-7}~\flux$ \citep[e.g.,][]{kajava2014,poutanen2014_4u1608}.
\end{tablenotes}
\end{threeparttable}
\end{table}

\subsection{Unsupervised spectral decomposition of the X-ray burst}\label{subsec:nmf}
Our spectral analysis presented in Section~\ref{subsec:spec} suggests that the accretion flux was enhanced during the X-ray burst. Since the intense X-ray burst emission may change the structure of the accretion disc or cool the corona \citep[e.g.,][]{maccarone2003,ballantyne2005}, it would not be surprising if the spectral shape of the persistent emission changed as well. However, the quality of the spectral data did not allow us to investigate this. 

Moving beyond conventional spectral fitting, unsupervised spectral decomposition methods have proven a powerful tool to map out the spectral variability of both supermassive and stellar-mass black holes \citep[][]{mittaz1990,vaughan2004,malzac2006,koljonen2013,parker2015}. For example, a study of the black hole LMXB GX 339--4 demonstrated how different emission components can be successfully disentangled and can reveal variations in the accretion disc in a regime where the disc flux is too low to be tracked by conventional spectral analysis \citep[][]{koljonen2015}. This is not unlike the situation for X-ray bursts, where the emission of the accretion flow is swamped by that of the burst. This motivated us to investigate the applicability of such techniques for X-ray burst studies. 

The aim of unsupervised decomposition methods is to find the minimum number of components that can together describe most of the physically interesting variability that is present in the data (in order words, it is an attempt to identify hidden structure in complex data). This can be achieved by using a mathematical approach in which the data are described as matrices. In this work we chose non-negative matrix factorisation \citep[NMF;][]{paatero1994,lee1999} as our spectral decomposition method, which was found to be most successful in disentangling different spectral components in the study of GX 339--4 \citep[][]{koljonen2015}. We briefly describe the NMF technique below for our specific application, but for general details the reader is referred to \citet{koljonen2015}. 

\subsubsection{NMF analysis}
The input for the NMF analysis is a data matrix composed of collection of discrete X-ray spectra \textsf{\textbf{X$_{ij}$}}, each consisting of counts detected in an energy bin $j$, measured at time $i$. Here we chose to divide the X-ray burst in equal time steps of 10 s to strike a balance between having a sizeable sample of spectra, but at the same time maintaining a sufficient number ($>$500) of counts per interval. This resulted in $i$$=$21 time-resolved spectra covering energies of 3--20 keV. Each of these spectra were then divided into $j$$=$25 energy slices with bin widths of 0.4--1.2~eV to ensure a positive flux in each energy bin. The NMF method then decomposes this input data matrix as $k$ separate source signals \textsf{\textbf{S$_{ki}$}} that are weighted over the energy bands as \textsf{\textbf{W$_{jk}$}}, such that the spectra are described as \textsf{\textbf{X}}$\approx$\textsf{\textbf{WS}}. The matrices \textsf{\textbf{S}} and \textsf{\textbf{W}} thus approximate the input collection of spectra \textsf{\textbf{X}}, and can be thought of as a spectral component (\textsf{\textbf{W}}, weights across energy bands) and its amplitude (\textsf{\textbf{S}}, signal). 

The source signal and the weight matrix that decompose the collection of input spectra are found by minimising a cost function under the constraint that they must be non-negative. To calculate the standard NMF \citep[][]{brunet2004}, we use the NMF package \citep[][]{gaujoux2010} that picks random starting values for \textsf{\textbf{S}} and \textsf{\textbf{W}} from a uniform distribution with values [0,max(\textsf{\textbf{X}})]. Using a multiplicative rule from \citet{lee2001}, the starting values are updated iteratively 10\,000 times to find a local minimum of the cost function. To avoid biases against local minima, we repeated the process for 300 different starting points. 

\begin{figure}
 \begin{center}
\includegraphics[width=8.5cm]{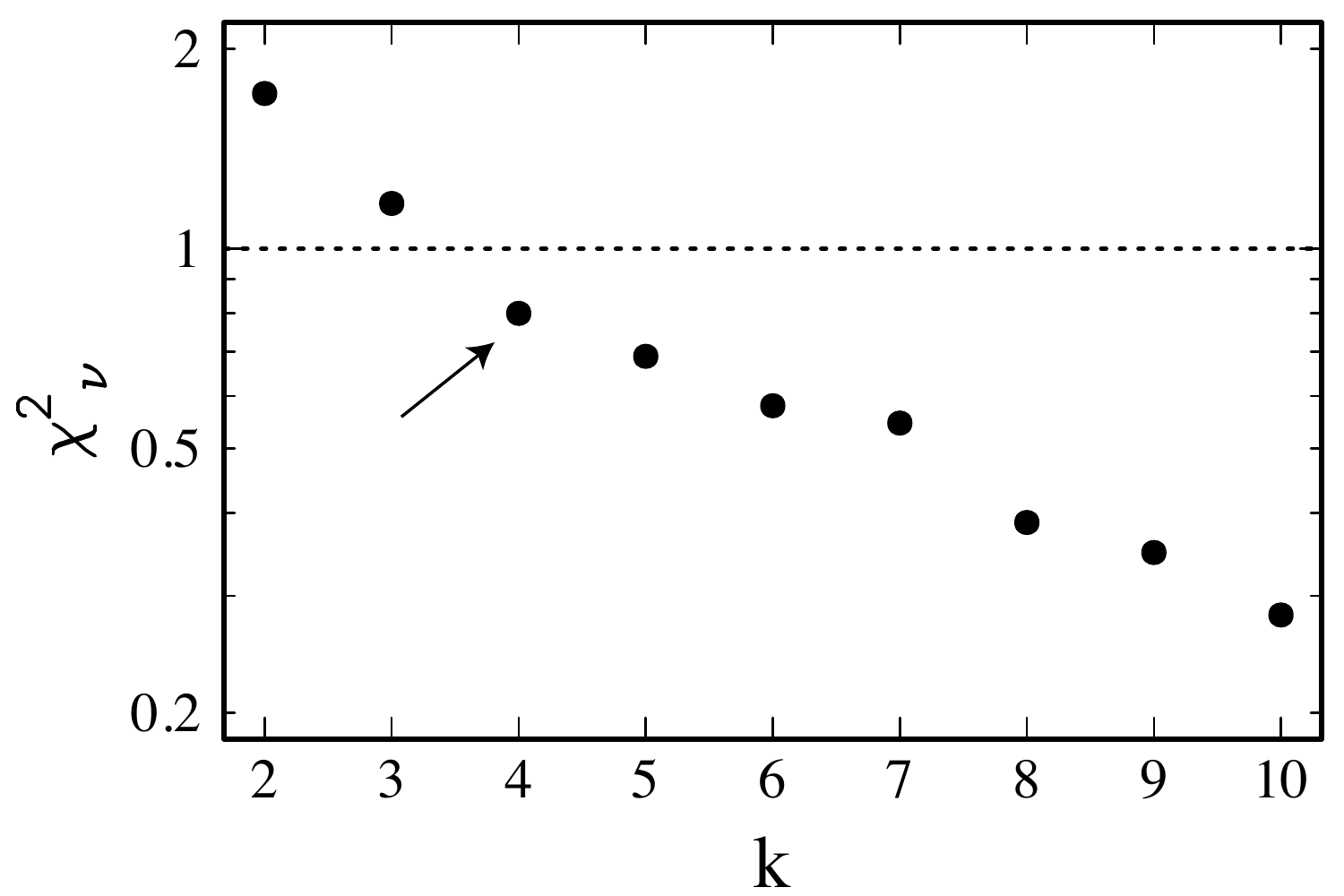} 
    \end{center}
    \vspace{-0.2cm}
\caption[]{{Quality measure of how well a certain factorisation ($k$) in the NMF analysis describes the data. This provides means to separate the significantly varying components from noise. We identify $k$$=$4 (see arrow) as the position where adding more components does not decrease $\chi^2$ strongly, and therefore chose that as the number of significant components.}}
 \label{fig:chi}
\end{figure}

The first step is to find the number of significant components $k$ that are needed to explain the variability in the data, i.e., the degree of factorisation. Statistical and systematic noise will give rise to components that are not related to physically interesting variations and therefore need to be filtered out. To achieve this, we use the $\chi^2$-diagram method outlined in \citet{koljonen2015}. This diagram provides a quality measure of how well a factorisation describes the data, and allows to reduce the number of components to only those ones that vary above the noise level. The choice for $k$ should be a knee in the diagram close to a value of $\chi^2=1$, where the slope changes from steep (increasing $k$ gives a large improvement) to more shallow (further increasing $k$ does not yield a large change in $\chi^2$ any more). Fig.~\ref{fig:chi} shows the $\chi^2$-diagram for our set of spectra. In this case we identify $k = 4$ as the position where adding more components does not decrease $\chi^2$ strongly, and therefore chose that as the number of significant components.

\begin{figure}
 \begin{center}
\includegraphics[width=8.5cm]{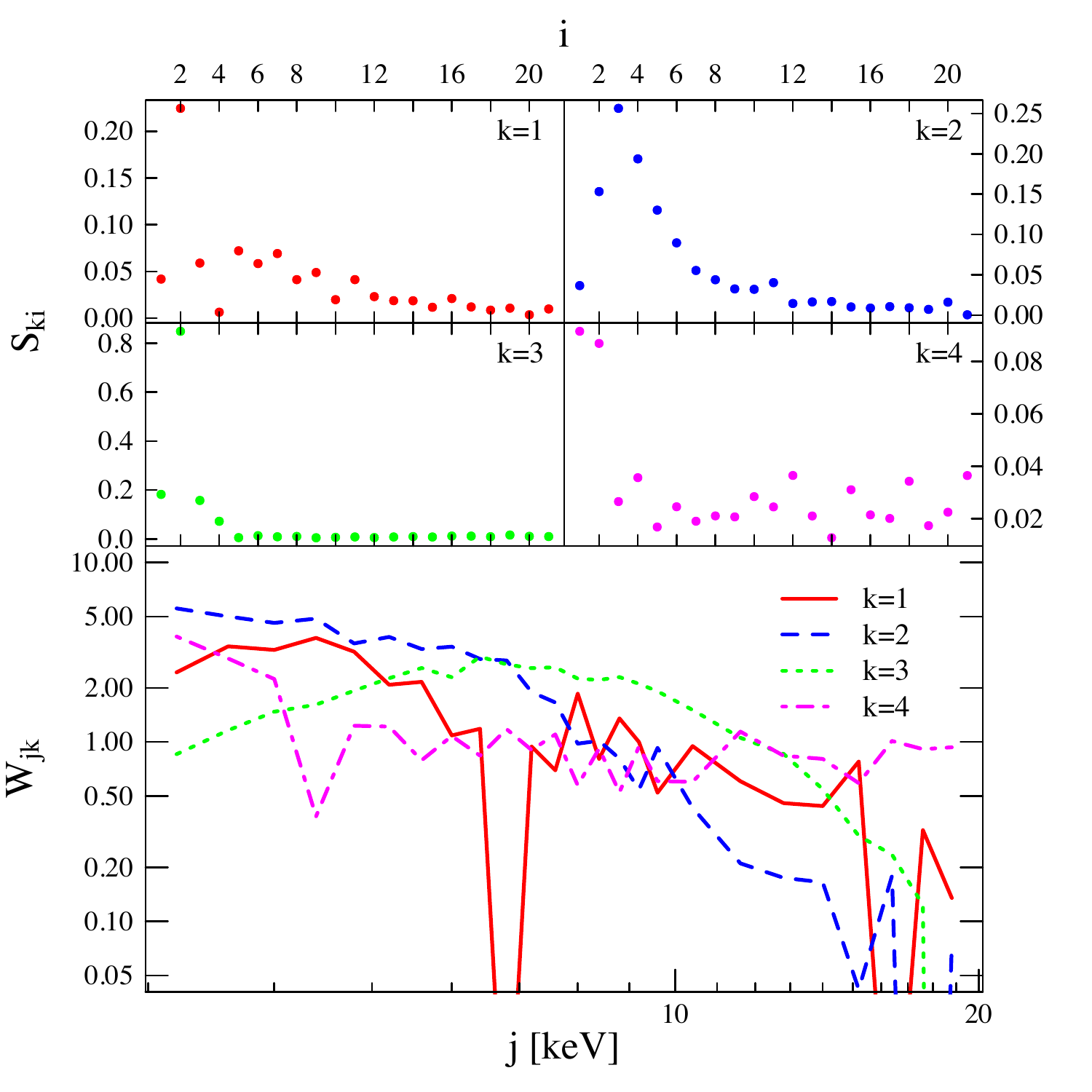}  
    \end{center}
        \vspace{-0.2cm}
\caption[]{{Results of the NMF analysis. The top four panels show how the significant components evolve over time (each time step $i$ represents 10 s), and lower panel shows their energy distribution (the energy slices $j$ have widths of 0.4--1.2 keV). Based on the strength of the source signals (\textsf{\textbf{S$_{ki}$}}) and the curvature of the energy spectra (\textsf{\textbf{W$_{jk}$}}) we identify the $k$=1,4 components with the accretion emission, and $k$=2,3 with the burst emission.}}
 \label{fig:nmf}
\end{figure}

We thus performed the NMF analysis assuming $k$=4 significant components, the results of which are shown in Fig.~\ref{fig:nmf}. The top four panels show how the amplitude of each NMF component changes over time, and their energy distribution is shown in the bottom panel. Note that the numbering of the components is random and does not reflect any particular order. The $k$=1 and $k$=4 components have a flat energy spectrum and a relatively weak signal, whereas the $k$=2 and $k$=3 components are characterised by a curved energy spectrum and a stronger signal. It is therefore likely that the $k$=1 and $k$=4 components are linked to the accretion emission, and the $k=$2 and $k=$3 components to the burst emission. The fact that there are two components for both the accretion flow and the burst emission, suggests that both do not only vary in normalisation but also in spectral shape. 

\subsubsection{Reconstruction of spectra from the NMF analysis}
The summed signals accurately describe the fluxes of the spectral components, but we do not expect a one-to-one correspondence between the individual NMF components and the spectral parameters (i.e., we cannot simply identify $k$=1 as the power-law index for instance). This is because the NMF is a linear technique and a varying power law index or blackbody temperature introduces non-linear effects \citep[e.g.,][]{mittaz1990}. However, we can assign physical meaning to the different emission components, and track the evolution of individual spectral parameters, by re-constructing spectra for each time interval from the NMF components as \textsf{\textbf{X$_{\mathrm{pow}}$}}$=\Sigma_{k=1,4}$\textsf{\textbf{W$_{jk}$}}$\times$\textsf{\textbf{S$_{ki}$}} and \textsf{\textbf{X$_{\mathrm{bb}}$}}$=\Sigma_{k=2,3}$\textsf{\textbf{W$_{jk}$}}$\times$\textsf{\textbf{S$_{ki}$}}. Fitting these NMF-built spectra (i.e., that have reduced dimensionality) in \textsc{XSpec} reveals that those constructed from the $k$=1,4 components can indeed be fitted by a power law with an index of $\Gamma$$\simeq$2, consistent with the pre-burst accretion emission. The spectra build from the NMF components $k$=2,3 can be fitted by a blackbody with a temperature of $kT_{\mathrm{bb}}$$\simeq$1--2~keV, as expected for an X-ray burst (Section~\ref{subsec:spec}).

Fig.~\ref{fig:nmfspec} shows our time-resolved analysis of the spectra constructed through the NMF. This sketches a similar picture as found from our conventional spectral analysis (Section~\ref{subsec:spec}): The burst emission (black histograms) peaks $\simeq$10~s after the initial rise, and can be described by a blackbody with a temperature varying between $kT_{\mathrm{bb}}$$\simeq$1.2 and 2.1~keV, and an inferred emission radius of $R_{\mathrm{bb}}$$\simeq$1--6~km (for $D$$=$3.6~kpc). This demonstrates that the NMF successfully reduced the data to its most significant components and is still able to accurately describe the evolution of the burst emission (i.e., retaining the most relevant information). In addition, the NMF allows us to resolve changes in the accretion emission (red histograms in Fig.~\ref{fig:nmfspec}). 

\subsubsection{Possible change in the accretion spectrum}
\begin{figure}
 \begin{center}
\includegraphics[width=8.0cm]{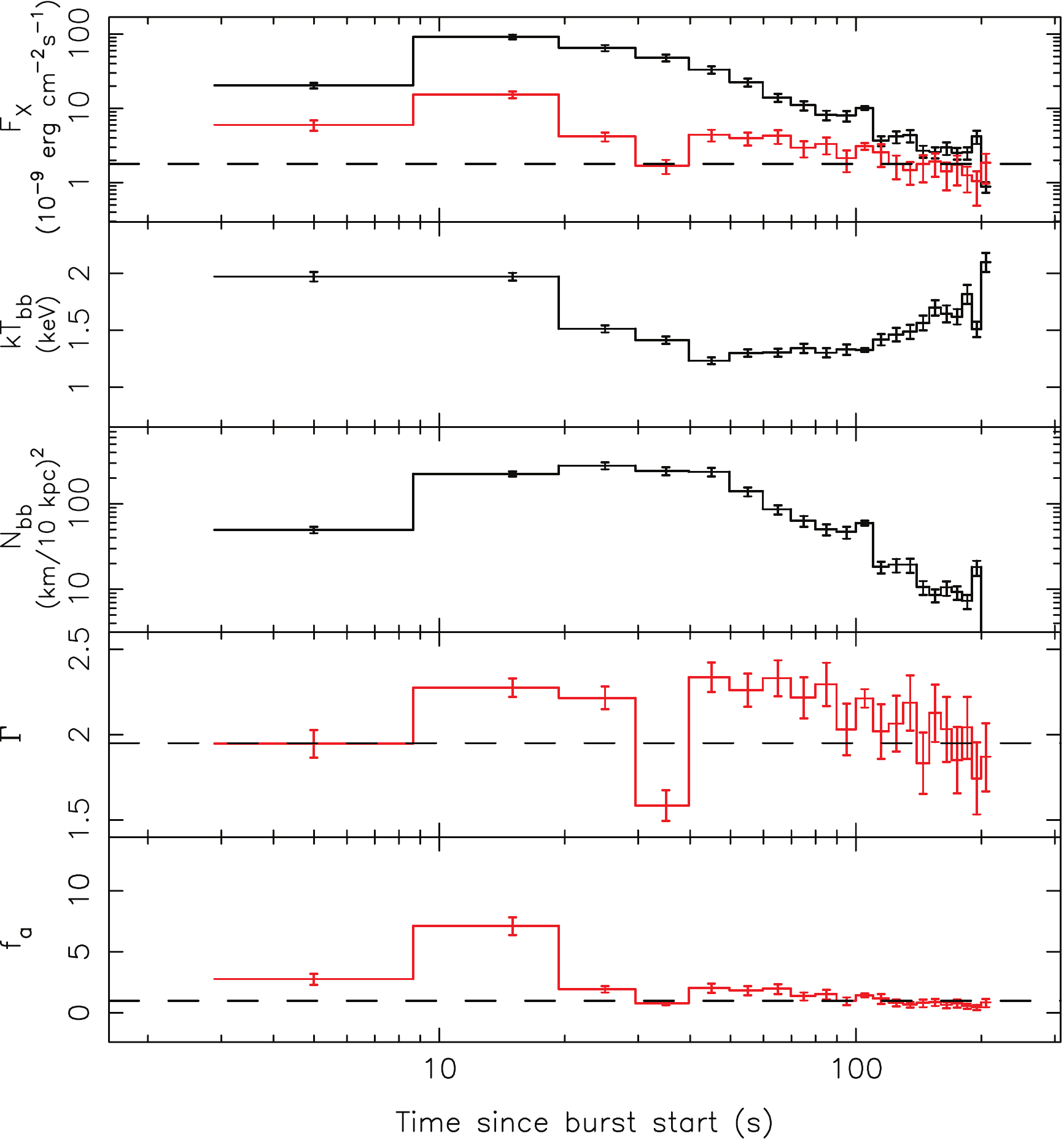} 
    \end{center}
    \vspace{-0.2cm}
\caption[]{{Analysis of the NMF-constructed spectra. Burst emission is indicated by black histograms, and accretion emission by red histograms. From top to bottom: the 3--20 keV flux, the blackbody temperature and normalisation, the power-law index, and the constant multiplication factor for the persistent flux $f_{a}$. In the plot for the power-law index the dashed horizontal line corresponds to the value inferred from the first interval. Errors represent 90 per cent confidence intervals here.}}
 \label{fig:nmfspec}
\end{figure}

As can be seen from the top panel in Fig.~\ref{fig:nmfspec}, the accretion flux appears enhanced compared to the pre-burst level (dashed line) until $\simeq$100~s into the burst. For comparison with our conventional spectral analysis, we divide the power-law flux during the burst by the pre-burst value ($F_{\mathrm{3-79}} = 2 \times 10^{-9}~\flux$; Section~\ref{subsec:spec}), to determine the constant multiplication factor $f_a$. This quantity is plotted in the bottom panel of Fig.~\ref{fig:nmfspec} and shows similar behaviour as inferred in Section~\ref{subsec:spec}: $f_a$ reaches a maximum value of several times the pre-burst level around the burst peak and returns to a value of $f_a$$\simeq$1 after $\simeq$100~s. The NMF analysis suggests that, in addition, the power-law slope is variable. It attends a value of $\Gamma$$\simeq$1.9 at the start (first 10-s spectrum) and end ($t$$>$100~s) of the X-ray burst, but is increased to $\Gamma$$\simeq$2.3 in between, except for a single low point with $\Gamma$$\simeq$1.6. This could indicate that the accretion emission softened in response to the X-ray burst. 

\begin{figure}
 \begin{center}
\includegraphics[width=8.0cm]{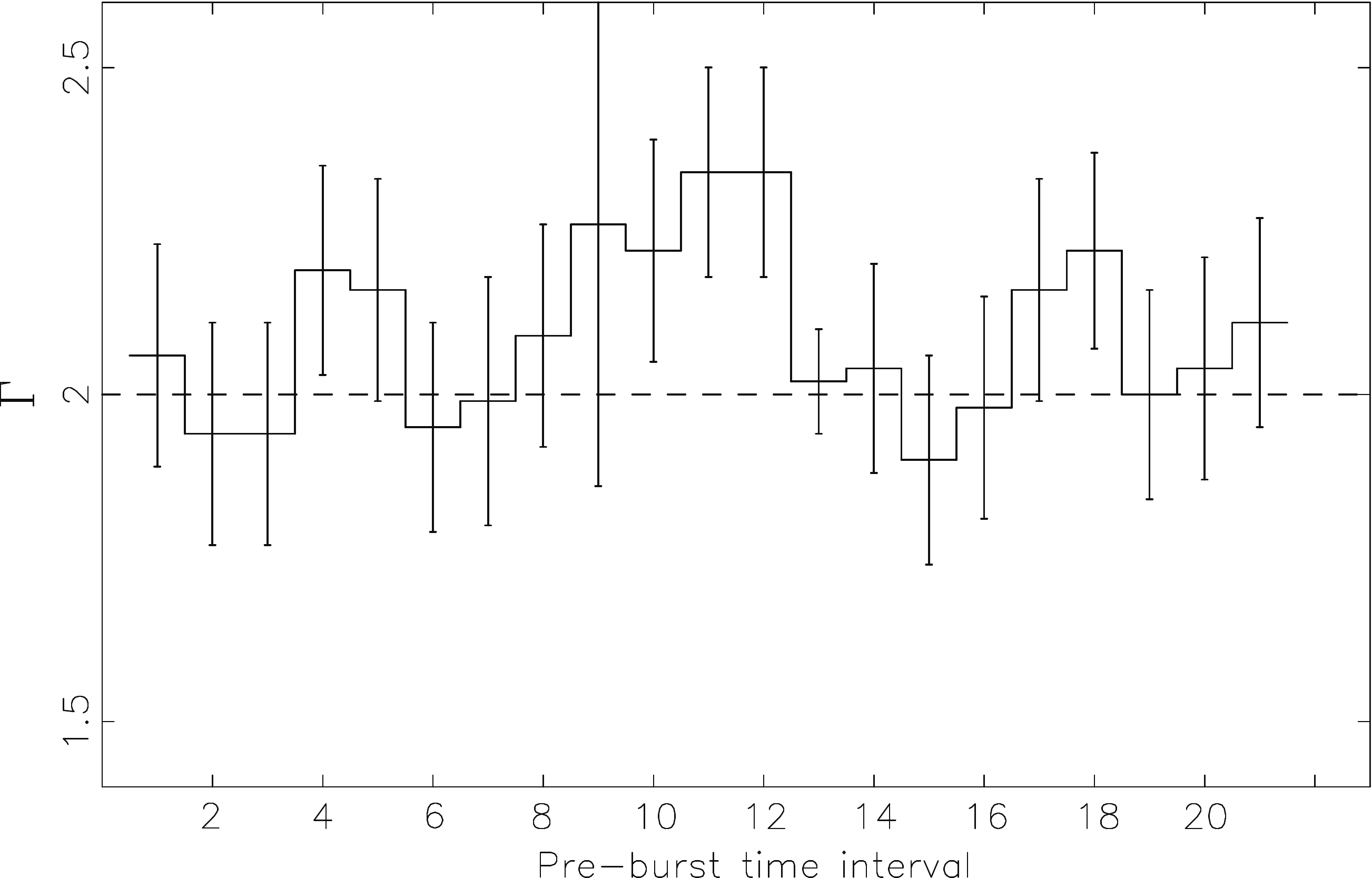} 
    \end{center}
    \vspace{-0.2cm}
\caption[]{{Power-law indices inferred from fitting 21 spectra of $\simeq$25-s length obtained from an interval of $\simeq$200--700~s prior to the X-ray burst. The spread in $\Gamma$ values is not unlike that seen in the fourth panel in Fig.~\ref{fig:nmfspec}, which indicates that the observed changes during the X-ray burst could be due to random variations. Errors are 90 per cent confidence intervals here. 
}}
 \label{fig:preburstspec}
\end{figure}

To investigate whether similar variability in the power-law slope is seen in the absence of a burst, we investigated a series of spectra extracted from an interval of $\simeq$200--700~s prior to the burst. We choose 25-s time steps to achieve an approximately similar number of counts per spectrum as for the burst spectra. The power-law indices inferred from fitting these 21 spectra with a simple power law (with $N_H$$=$$1.5\times10^{22}~\nh$ fixed) is shown in Fig.~\ref{fig:preburstspec}. The index varied irregularly between a minimum/maximum value of $\Gamma=1.9/2.3$ with typical errors of $\pm0.2$ (90 per cent confidence). This is not unlike the variability seen during the X-ray burst. Therefore we cannot conclusively ascribe the apparent softening as being due to the X-ray burst rather than random variations.

%%%%%%%%%%%
% DISCUSSION
%%%%%%%%%%%

\section{Discussion}\label{sec:discussion}
We report on the detection of a type-I X-ray burst during a \nustar\ observation of the neutron star LMXB \source. The burst had a
duration of $t_{\mathrm{burst}}$$\simeq$200~s, a relatively low peak flux of $F_{\mathrm{peak}}$$\simeq$0.3--0.5$~F_{\mathrm{Edd}}$, and a total radiated energy of $E_{\mathrm{rad}}$$\simeq$$1\times 10^{38}~$erg. Detailed analysis of the non-burst emission revealed that the burst occurred during a hard X-ray spectral state, with the neutron star accreting at a few per cent of the Eddington limit \citep[][]{degenaar2015_4u1608}. 

\vspace{-0.4cm}
\subsection{Enhancement of the persistent emission}\label{subsec:persenhance}
The fact that \source\ was in a hard X-ray spectral state made it possible to address whether the accretion emission was altered by the intense X-ray burst emission. We performed time-resolved spectroscopy of the burst and the underlying accretion emission, where we allowed the latter to change in flux. Despite that the burst was not very long, bright, or energetic, we found that the accretion flux was enhanced by a factor of $\simeq$5 during the peak of the X-ray burst and returned to its pre-burst level after $\simeq$100~s. Such enhancements have been detected in a large number of bursts from many different sources, both during hard and soft X-ray spectral states, and both for PRE and non-PRE bursts \citep[][]{intzand2013,worpel2013,worpel2015,ji2014_4u1608,ji2015_gs1862,peille2014}. 

The burst-accretion interaction is complex, but the increase of the persistent emission during bursts has been associated with the Poynting-Robertson drag, in which transfer of momentum by the burst photons reduces the angular momentum of the inner accretion disc, hence allowing more matter to fall towards the neutron star and enhance the accretion rate \citep[e.g.,][]{walker1992}. For \source, analysis of the reflection spectrum revealed that the inner accretion disc resided close to the neutron star \citep[$R_{\mathrm{in}}$$\simeq$7--10~$\gmc$;][]{degenaar2015_4u1608}, so this seems a viable explanation for the observed enhancement in accretion flux. We note that studies of the rapid time variability in 4U 1608--52 suggested (small) changes in the inner accretion disc radius in response to nuclear burning on the neutron star surface \citep[][]{yu2002}, which may fit in with this picture. Reprocessing of burst photons by the accretion disc or corona could possibly also enhance the accretion emission during an X-ray burst \citep[e.g.,][]{ballantyne2005,intzand2013}.

Apart from changes in the accretion flux {\it during} the X-ray burst, we observed a small ($\simeq$3 per cent) increase in intensity that started $\simeq$0.5~h \textit{prior} to the X-ray burst detection. Such a rise was not seen during any of the other \nustar\ orbits, which instead showed the count rate to gradually decrease over time. An enhanced accretion flux was also reported prior to three superbursts \citep[][]{kuulkers2002,chenevez2011_superburst,keek2014}. For two of these, the accretion emission remained enhanced directly after a normal X-ray burst was ignited, $\simeq$0.5~h before the superburst. The enhanced emission could therefore be related to the neutron star surface remaining hot after the normal burst. For \source, there was a data gap of $\simeq$2.3~ks before the satellite orbit in which the enhancement was observed, so a prior X-ray burst could have been missed. However, it seems that the accretion emission was not enhanced right from the start of this orbit, but began to rise only $\simeq$1~ks into the observation. Arguably, an association with a missed X-ray burst therefore seems unlikely.

\vspace{-0.4cm}
\subsection{Possible change in the shape of the accretion spectrum}\label{subsec:shapechange}
The intense radiation from an X-ray burst may clear, or change the shape of, the inner accretion disc \citep[see e.g.,][for a discussion of various possible effects]{ballantyne2005}. Furthermore, the presence of a corona (a hot population of electrons) can cause burst photons to Compton up-scatter and hence produce a high-energy emission tail. In turn, the burst photons may be able to efficiently cool or even eject the corona \citep[e.g.,][]{maccarone2003,kluzniak2013,ji2015_gs1862}. Therefore, it seems reasonable to expect that an X-ray burst does not only change the flux, but also the spectral shape of the accretion emission.

Conventional time-resolved spectroscopy of the burst did not allow us to address whether the shape of the accretion emission was changing along with its flux. However, we were able to address this in more detail by using an unsupervised spectral decomposition method, in which the data are reduced to only the most significantly varying components \citep[e.g.,][]{koljonen2015}. This way we found that the accretion emission possibly softened during the X-ray burst. However, investigating short ($\simeq$25~s) time slices of the accretion emission prior to the X-ray burst showed changes of similar magnitude. Therefore, we cannot exclude random variations as the cause of the apparent softening. Nevertheless, it is suggestive that the power-law index was increased during several consecutive intervals around the burst peak and returned to the pre-burst level at the same time as the accretion flux settled back to its pre-burst value. Therefore, it is interesting to explore the physical interpretation of the possible softening of the accretion emission.

\citet{keek2014} reported changes in the spectral shape of the accretion emission during a (non-PRE) superburst from 4U 1636--536. It was found that the persistent emission increased by a factor of $\simeq$1.8, and that the cut-off energy decreased from $\simeq$20 to 10~keV during the X-ray burst. This may be interpreted as cooling of the corona by the injection of the soft burst photons \citep[][]{keek2014}. In case of \source, there was no detectable high-energy cutoff in the non-burst accretion emission \citep[][]{degenaar2015_4u1608}. However, the possible softening of the power-law index might indicate the same physical process.

Cooling of the corona has also been invoked to explain the hard X-ray ($>$30~keV) deficits seen when stacking hard state bursts of a number of sources \citep[][]{chen2012_xrbs_igrj1747,chen2013_aqlx1_bursts,ji2013_xrbs_4u1636,ji2014_bursts,ji2014_aa}. In case of \source, we did not find such a deficit in the 30--79~keV light curve during the \nustar-detected X-ray burst. A stacking study of a large sample of X-ray bursts detected from \source\ with \rxte\ did not reveal a hard X-ray deficit either \citep[][]{ji2014_aa}. Such a deficit may be absent if the corona is far away, or if the hard X-ray source is in fact an equatorial boundary/spreading layer on the neutron star surface \citep[e.g.,][]{kluzniak1991,inogamov1999,popham2001}, so that most of the burst flux does not pass through the ``corona'' and therefore does not result in notable cooling. Detailed analysis of the \nustar\ reflection spectrum revealed a compact corona with a height of only $h$$<$8.5~$\gmc$ above the neutron star, which may favour the boundary layer scenario. However, given the low count rate at $\gtrsim$30~keV, sensitivity limits may have prevented us from detecting a high-energy deficit during the burst.

Since the increase in power-law index (i.e., possibly representing cooling) persisted as long as the accretion emission was enhanced, one may wonder whether cooling of the corona due to an X-ray burst could condense matter back on the disc and thereby temporarily enhance the accretion rate. We can roughly estimate the mass-accretion rate on to the neutron star in \source\ from the average persistent flux measured by \nustar\ ($F_{\mathrm{3-79}}$$\simeq$$2 \times 10^{-9}~\flux$). Assuming a bolometric correction factor of $\simeq$2, this suggests $L_{\mathrm{bol}}$$\simeq$$6\times 10^{36}~\dist~\lum$ and $\dot{M}$=$LR/GM$$\simeq$$3\times10^{16}~\mdotgs$ (where $G$ is the gravitational constant and $M$$=$$1.4~\Msun$ and $R$$=$$10$~km are the adopted mass and radius of the neutron star, respectively). If we take the accretion rate to be enhanced by a factor $\simeq$2 on average for a duration of $\simeq$100~s, this suggests that $\simeq$$7\times10^{18}$~g ($\simeq$$3\times10^{-15}~\Msun$) of additional mass was accreted in response to the X-ray burst. 

If the corona is spherical with a radius of $\simeq$10~km and a surface density of $\simeq$$10^{24}~\mathrm{g~cm}^{-2}$ (corresponding to the Thomson optical depth of the corona of $\tau$$\simeq$1), this would imply a total number of $\simeq$$3\times10^{36}$~particles or $\simeq$$5\times10^{12}$~g (for a proton mass of $1.67\times10^{-24}$~g). This falls short by orders of magnitude of our simple estimate above. Therefore, there was likely not enough mass in the corona to account for (all of) the additional matter that was accreted during the X-ray burst.

We note that the above discussion is based on the assumption that the burst emission can be well-approximated by a pure blackbody. If the spectrum is harder, as may be expected due to atmospheric effects \citep[e.g.,][]{foster1986}, we might see the persistent emission compensating for this in our model fits by getting softer. However, deviations from a blackbody are expected to be strongest near the Eddington flux \citep[e.g.,][]{suleimanov2011}, whereas this burst peaked below $\simeq$50 per cent of the Eddington limit. Realistic atmosphere models for X-ray bursts are complex, hence complicating spectral fitting, and do not yield better fits to the data than a blackbody \citep[e.g.,][]{intzand2013,keek2014}. 

\vspace{-0.2cm}
\subsection{The prospects of \nustar\ and NMF for burst studies}\label{subsec:prospects}
Due to the hard X-ray spectrum of \source, \nustar\ proved very valuable to disentangle the accretion and burst emission. Moreover, a strong aspect of \nustar\ is that it provides excellent reflection spectra that can place strong constraints on the accretion geometry in X-ray binaries, such as the location of the inner accretion disc and the corona \citep[e.g.,][]{fuerst2015,miller2015,parker2015}. Analysis of the \nustar\ reflection spectrum of \source\ was presented in \citet{degenaar2015_4u1608}. As discussed in Section~\ref{subsec:shapechange}, these tight constraints on the accretion geometry  provide valuable information for interpreting the effect of X-ray bursts on the accretion emission.

The quality of current data does not allow us to detect changes in the accretion emission during bursts via spectral analysis \citep[see also][]{worpel2015}, except in special cases \citep[][]{keek2014}. Applying unsupervised spectral decomposition methods can possibly provide more insight into this. Our NMF analysis of the burst of \source\ proved that the method was successful in disentangling the accretion emission from that of the burst. The evolution of the burst emission inferred from the NMF yielded similar results to that found in conventional spectral fitting. In addition, it allowed to better resolve changes in the accretion emission. Applying NMF or similar methods to a larger sample of X-ray bursts may be a promising avenue to gain further insight into the burst-accretion interaction, even before the launch of future X-ray missions.

\vspace{-0.5cm}
\section*{Acknowledgements}
ND acknowledges support via an EU Marie Curie Intra-European fellowship under contract no. FP-PEOPLE-2013-IEF-627148.  DA is supported by the Royal Society. This work is based on data from the \nustar\ mission, a project led by California Institute of Technology, managed by the JPL, and funded by NASA. We made use of \maxi\ data provided by RIKEN, JAXA and the MAXI team, and publicly available light curves from the \swift/BAT transient project. The authors are grateful to the anonymous referee for valuable comments that helped improve this manuscript.

\vspace{-0.5cm}
\footnotesize{

}

\end{document}